# Soliton eigenvalue control with optical lattices

Yaroslav V. Kartashov,[1,2] Lucian-Cornel Crasovan,[1,3] Anna S. Zelenina,[2] Victor A. Vysloukh,[4] Anna Sanpera,[1,5] Maciej Lewenstein,[1,5] and Lluis Torner[1]

[1]*ICFO-Institut de Ciencies Fotoniques, and Department of Signal Theory and Communications, Universitat Politecnica de Catalunya, 08034 Barcelona, Spain*
[2]*Physics Department, M. V. Lomonosov Moscow State University, 119899 Moscow, Russia*
[3]*Department of Theoretical Physics, Institute of Atomic Physics, PO Box MG-6, Bucharest, Romania*
[4]*Departamento de Fisica y Matematicas, Universidad de las Americas – Puebla, CP 72820, Puebla, Cholula, Mexico*
[5]*Institut für Theoretische Physik, 30167 Hannover, Germany*

We address the dynamics of higher-order solitons in optical lattices, and predict their self-splitting into the set of their single-soliton constituents. The splitting is induced by the potential introduced by the lattice, together with the imprinting of a phase tilt onto the initial multisoliton states. The phenomenon allows the controllable generation of several coherent solitons linked via their Zakharov-Shabat eigenvalues. Application of the scheme to the generation of correlated matter waves in Bose-Einstein condensates is discussed.

During the last decades the concept of soliton has penetrated almost all areas of physics including hydrodynamics, plasma physics, optics [1], and, recently, Bose-Einstein condensates (BECs) in cold gases [2,3]. Solitons are formed when the linear effects that cause spreading of wave packets are balanced properly by nonlinearity. In the case of single solitons, such balance is remarkably robust, a property that makes them suitable for the transmission and manipulation of, e.g., light and matter. In particular, optical solitons have been thoroughly studied due to their potential applications in telecommunications. In soliton-based communication systems each soliton can be used as a bit of information, but methods to encode and to manipulate information in higher-order, or multisoliton bound states (BSs) have been also proposed, e.g., for security enhanced information transmission [4]. In physical systems modeled by so-called completely integrable evolution equations, such multisoliton states are made of sets of several individual solitons, with different amplitudes, glued together with zero binding energy. The amplitudes of the solitons "hidden" inside the BS are given by the corresponding Zakharov-Shabat (ZS) eigenvalues [1,4]. Because there is no binding energy between the solitons forming the BSs, suitable perturbations can be used to split them into their constituents [5].

In this Letter we propose to use weak optical lattices to induce the self-splitting of BSs made of either optical or mater waves, and thereafter to control the velocities of the product single solitons. The optical lattices – periodic light patterns that act like multiple potential wells – has been demonstrated in nonlinear optics [6-8] and in BECs [9]. They can be also used for particle sorting [10] or for trapping arrays of neutral atoms [11]. Solitons in lattices have been studied extensively in the discrete, or tight-binding limit, in nonlinear optics (see [12]) and are starting to be investigated in BECs [13,14]. An important feature of the condensates is the possibility to continuously tune their nonlinearity, which is proportional to the scattering length $a_s$ that characterizes two-body interactions, by using Feschbach resonances [15]. With this technique the concept we put forward here can be tested in condensates.

The generic equation describing the evolution of optical (matter) wave packets in the presence of Kerr (mean field cubic) nonlinearity and a periodic potential induced by a weak optical lattice is the nonlinear Schrödinger equation:

$$i\frac{\partial \Psi}{\partial \xi} = -\frac{1}{2}\frac{\partial^2 \Psi}{\partial \eta^2} + \gamma |\Psi|^2 \Psi - pR(\eta)\Psi. \qquad (1)$$

In the optical context $\Psi$ is the wave function, $\xi$ and $\eta$ are the longitudinal and transverse coordinates, respectively, $\gamma$ is the strength of the cubic nonlinearity, $p$ is the potential depth and $R(\eta)$ is the potential profile. Throughout this paper we assumed a periodic potential $R(\eta) = \cos(2\pi\eta/T)$, with the period $T$. Eq. (1) models laser beam propagation in slab waveguides. In this case $\Psi$ is proportional to the slowly varying envelope of light field, the longitudinal coordinate $\xi$ is scaled in diffraction lengths $L_{\text{dif}} = kr_0^2$, with $r_0$ being the beam width. The transverse coordinate $\eta$ is expressed in units of $r_0$, whereas the lattice depth is given by $p = L_{\text{dif}} / L_{\text{ref}}$, where $L_{\text{ref}} = c/(\delta n\,\omega)$ and $\delta n$ is the modulation depth of refractive index. For a Kerr self-focusing medium $\gamma = -1$, whereas for a defocusing one $\gamma = 1$.

In the case of matter waves Eq. (1) describes dynamics of a one-dimensional Bose-Einstein condensate confined in an optical lattice generated by means of a standing laser wave of wavelength $\lambda$. Now variable $\xi$ stands for time in units of $\tau = 2m\lambda^2/\pi h$, with $m$ being the mass of the atoms and $h$ the Planck's constant, $\eta$ is the longitudinal coordinate along the axis of the quasi-one-dimensional condensate expressed in units of $\lambda \pi^{-1}$. For typical experiments $\lambda$ ranges from 0.8 to 3.2 $\mu$m. Parameter $p$ is proportional to lattice depth $E_0$ expressed in units of the recoil energy $E_{\text{rec}} = h^2/2m\lambda^2$. Lattice depths of the order $E_0 \leq 22 E_{\text{rec}}$ ($p \leq 22$) have already been achieved [9]. In the quasi-one-dimensional BEC $\gamma = 2\lambda a_s N_a/\pi \ell^2$, where $a_s$ is the s-wave scattering length, $N_a$ is the number of atoms, and $\ell$ is the harmonic oscillator length in the transverse direction [16]. The sign and value of $a_s$ and thus of $\gamma$ can be changed by varying the applied magnetic field [15]. Positive $\gamma$ stands for repulsive interactions while negative $\gamma$ for attractive ones. Here we assume attractive, or self-focusing, interactions with $\gamma = -1$, and vary the depth of the periodic potential from small values $p \sim 0.05$, when the potential can be treated as a small perturbation, to values $p \sim 1$, when the potential is of the order of nonlinearity. Also, we set $T = \pi/2$, but the result can be extended to other values of the period by using the scaling properties of the evolution Eq. (1).

We aim at the possibility of extracting and controlling the dynamics of the single-soliton constituents of the $N$-soliton bound state corresponding to the initial conditions: $\Psi(\eta,\xi=0) = N\,\text{sech}(\eta)\exp(i\alpha_{\text{in}}\eta)$, where $\alpha_{\text{in}}$ is the phase tilt or angle. Here we will restrict ourselves to BSs with $N = 2,3$, thus carrying 2 or 3 single solitons. The amplitudes $\chi_k$ and angles $\alpha_k$ of the constituent solitons in BS are related with ZS eigenvalues $\lambda_k$ as $\chi_k = 2\,\text{Im}\,\lambda_k$ and $\alpha_k = 2\,\text{Re}\,\lambda_k$ [1,4]. Thus for BS $\Psi(\eta,\xi=0) = N\,\text{sech}(\eta)\exp(i\alpha_{\text{in}}\eta)$ the amplitude and angle of the $k$-th constituent soliton are given by $\chi_k = 2k-1$ and $\alpha_k = \alpha_{\text{in}}$ ($k = 1,...,N$), respectively. In the absence of perturbations such BS exhibits periodic breathing and recovering of its initial shape upon propagation. In the context of optical solitons different types of perturbations, such as third-order group-velocity dispersion, stimulated Raman scattering, two photon absorption, cascading, or asymmetric spectral filtering, are known to lead to the splitting of the BSs into the constituent solitons, so the ZS eigenvalues can be extracted BS [5]. The method we put forward here, based on the use of an optical lattice, is extremely robust and controllable. We show that the splitting can be controlled by acting on two external parameters, namely a phase tilt imposed on the BS and, more importantly, the lattice strength.

In Fig. 1 we show a few representative examples of the typical decay of three-soliton bound states induced by the periodic potential produced by the optical lattice. The plots have been obtained for the weak potential $p = 0.05$. As shown in Fig. 1(a) and 1(b), for small values of $p$ and small incident angles $\alpha_{\text{in}}$ only the soliton with highest energy is trapped inside the central waveguide created by the lattice sites, whereas the other ones move across the periodic potential at constant angles. We found that the trapping occurs only when the initial tilt is below a critical value. For single solitons, the value of the critical tilt can be estimated analytically, to find [6,8]

$$\alpha_{\text{cr}} = 2\left[\frac{p\pi^2}{T\chi}\sinh^{-1}\left(\frac{\pi^2}{T\chi}\right)\right]^{1/2}, \qquad (2)$$

This estimate was found to agree very well with the numerical values (e.g., up to the 4th digit for $\chi = 5$, $p = 0.05$). Importantly, for weak lattices the amplitudes $\chi_k$ of the output solitons arising after the splitting were always found to almost coincide with the initial values hidden inside the BS. When the outgoing solitons propagate across the lattice, their amplitudes decrease slowly because of the soliton interaction and crossing of the corresponding periodic potential. Detailed estimates of such small radiation are given in Refs. [6]. Physically, the splitting is caused by the different phase-shifts acquired by the several solitons contained in the unperturbed BS through their different scattering by the lattice and, more importantly, by their mutual nonlinear interactions. In the early stages of the evolution ($\xi \sim 1$), the tilted three-soliton BS starts a self-compression, followed by reshaping into a twin-peak structure. Asymmetrical energy tunnelling to the neighbouring lattice sites finally leads to the splitting and soliton escaping. Complex interactions between the solitons may lead to repulsion, hence solitons bounce back against the initial tilt, as visible in Fig. 1.

Fig. 1 shows that the escape angles of the solitons can be effectively controlled by varying the tilt $\alpha_{\text{in}}$. To further explore this point we plotted in Fig. 2 the escape angles $\alpha_k$ versus the initial tilt $\alpha_{\text{in}}$ for two- and three-soliton BSs. The critical tilt beyond which the highest energy soliton escapes from the central potential well is close to 0.4 for $N = 2$ and 0.45 for $N = 3$. Large tilts make solitons moving closer to each other whereas small ones correspond to an ideal regime for soliton switching and steering. Importantly, we found that the splitting process is not sensitive against variations in the lattice strength or against small random noise present in the complex profile of the input BS (see Figs. 2(c) and 2(d)). This result claims for the remarkable robustness of the eigenvalue control process afforded by the lattice, as well as for the feasibility of its experimental observation.

Another important possibility opened by the lattice is tuning its strength, thus a new control parameter to the toolkit aimed at controlling the soliton splitting. This possibility is illustrated in Fig. 3(a) and Fig. 3(b), which show the splitting of BSs at fixed initial tilt by varying the lattice strength. As one can see from the plots, large variations of the potential depth do not lead to qualitative changes in the splitting dynamics. Fig. 3(c) summarizes the point: The escape angles of the solitons generated through the splitting growth smoothly with the lattice strength. It is remarkable that trajectories of escaping solitons remain almost linear up to rather high values of the potential depth $p \sim 0.3$.

As it is clear from the previous discussion, on one hand the periodic potential induces the break-up of the BSs into its constituents. However, on the other hand it also provides a potential barrier for each soliton. Therefore, when such a potential barrier is high enough it might lead to trapping of the generated single solitons into a specific channel of the lattice, thus providing a method to harvest the product solitons for further manipulation. To explore this phenomenon quantitatively, let a soliton be trapped in the $n$-th lattice channel when the coordinate $\eta_\mathrm{m}$ of its intensity maxima satisfies $nT - T/2 \leq \eta_\mathrm{m} \leq nT + T/2$ at $\xi \to \infty$. In Fig. 4 we present the outcome of an illustrative simulation for two different BS decay scenarios in the deep potential $p \sim 1$. In this regime, the decay of BS and the subsequent propagation of the produced single solitons are accompanied by the generation of radiation. The generated solitons might be trapped in different lattice channels depending on the value of the initial tilt. However, in this regime the process is highly inelastic, thus the amplitudes of the output solitons may depart considerably from the values $\chi_k = 2k - 1$ carried by the input BS. As in the case of small potential depth, the soliton with the highest energy can no longer be trapped in the central channel when incident angle exceeds a threshold value. Naturally, large tilts lead to large radiative losses that affect drastically the soliton amplitudes and make the trapping impossible. Also, for deep enough potentials the annihilation or birth of new solitons might occur. Radiative losses grow dramatically when the tilt angle approaches the Bragg one, so resonances with spectral bands become important [6]. However such angles fall far above those considered here. A detailed description of the Bragg scattering of wavepackets in periodic potentials in linear and nonlinear regimes can be found in Ref. [17], where it is shown that scattering occurs in accordance with the contributions of different Bloch states into the spectrum of the wavepacket. The approach of Ref. [17] can potentially be applied to scattering of BSs since those can be viewed as nonlinear combination of several single solitons (wavepackets).

A central motivation of this paper is the implementation of the proposed scheme in one-dimensional Bose-Einstein condensates hold in optical lattices, thus next we discuss the feasibility of the method with currently available technology. Creation of a multi-soliton bound state can be achieved in the following way. First, a single standing bright soliton condensate has to be created in the absence of the optical lattice. Such soliton corresponds to the ground state of the system for a certain (negative) value of the scattering length. For instance, for a scattering length such that $\gamma = -1/4$ the profile of such soliton is given by $\Psi(\eta) = 2\,\text{sech}(\eta)$. By changing then the scattering length to $\gamma = -1$ with the aid of Feschbach resonances, $\Psi(\eta)$ becomes a bound state of two solitons. The optical lattice is then grown up adiabatically. Using a phase-imprinting technique [18], or by tilting the optical lattice in the gravitational field, one can imprint onto the condensate wave function the desired linear phase. All these steps are currently standard experimental techniques. Notice that the creation of higher-order solitons in BECs is discussed also in Ref [19].

To summarize, we have proposed a mechanism for the extraction and control of the multiple single solitons carried by higher-amplitude initial conditions in systems modelled by the nonlinear Schrödinger equations with external linear periodic potentials. We discussed the physical implementation of the technique in optical Kerr media and in Bose-Einstein condensates hold in optical lattices. For weak lattices, we showed that the lattice strength and a phase tilt imprinted to the initial conditions enable extraction of all the ZS eigenvalues carried by the input conditions. Such splitting process being coherent, it leads to the generation of quantum correlated matter waves, a feature that might find applications in the area of macroscopic coherent atomic ensembles [20].

This work has been partially supported by the Generalitat de Catalunya, the Spanish Government through grant BFM2002-2861, and the Deutsche Forchungsgemeinschaft (SFB-407, SPP1116).

# References

1. N. N. Akhmediev and A. Ankiewicz, *Solitons: Nonlinear Pulses and Beams* (Chapman & Hall, London, 1997); Y. S. Kivshar and G. P. Agrawal, *Optical Solitons: From Fibers to Photonic Crystals* (Academic Press, San Diego, 2003).
2. S. Burger *et al.*, Phys. Rev. Lett. **83**, 5198 (1999); J. Denschlag *et al.*, Science **287**, 97 (2000).
3. L. Khaykovich *et al.*, Science **296**, 1290 (2002); K. E. Strecker *et al.*, Nature (London) **417**, 150 (2002).
4. A. Hasegawa and T. Nyu, J. Lightwave Technol. **11**, 395 (1993).
5. P. K. A. Wai *et al.*, Opt. Lett. **11**, 464 (1986); K. Tai *et al.*, Opt. Lett. **13**, 392 (1988); V. V. Afanasyev *et al.,* Opt. Lett. **15**, 489 (1990); Y. Silberberg, Opt. Lett. **15**, 1005 (1990); S. R. Friberg and K. W. DeLong, Opt. Lett. **17**, 979 (1992); L. Torner *et al.,* Opt. Lett. **21**, 462 (1996); M. Golles *et al.,* Phys. Lett. A **231**, 195 (1997); V. A. Aleshkevich *et al.*, Quantum Electron. **33**, 460 (2003); K. S. Lee and J. A. Buck, J. Opt. Soc. Am. B **20**, 514 (2003); V. A. Aleshkevich *et al.*, Opt. Lett. **29**, 483 (2004).
6. R. Scharf and A. Bishop, Phys. Rev. A **46**, R2973 (1992); Phys. Rev. E **47**, 1375 (1992).
7. N. K. Efremidis *et al.*, Phys. Rev. E **66**, 046602 (2002); J. W. Fleischer *et al.*, Nature (London) **422**, 147 (2003).
8. D. Neshev *et al.*, Opt. Lett. **26**, 710 (2003); P. J. Y. Louis *et al.*, Phys. Rev. A **67**, 013602 (2003); N. K. Efremidis and D. N. Christodoulides, Phys. Rev. A **67**, 063608 (2003); Y. V. Kartashov et al., Opt. Lett. **29**, 766 (2004); **29**, 1102 (2004); Opt. Exp. **12**, 2831 (2004).
9. A. Anderson and M. A. Kasevich, Science **282**, 1686 (1998); F. S. Cataliotti *et al.*, Science **293**, 843 (2001); M. Greiner *et al.*, Phys. Rev. Lett. **87**, 160405 (2001).
10. M. MacDonald, G. Spalding, and K. Dholakia, Nature (London) **426**, 421 (2003).
11. G. K. Brennen *et al.*, Phys. Rev. Lett. **82**, 1060 (1999).
12. D. N. Christodoulides, F. Lederer, and Y. Silberberg, Nature (London) **424**, 817 (2003).

13. A. Trombettoni and A. Smerzi, Phys. Rev. Lett. **86**, 2353 (2001); F. K. Abdullaev *et al.*, Phys. Rev. A **64**, 043606 (2001).

14. B. Eiermann *et al.*, Phys. Rev. Lett. **92**, 230401 (2004).

15. S. Inouye *et al.,* Nature (London) **392**, 151 (1998), S. Cornish *et al.,* Phys. Rev. Lett. **85**, 1795 (2000).

16. For $\lambda = 1\,\mu$m and $\ell \approx 0.5\,\mu$m the value $\gamma \approx 1$ can be achieved in $^6$Li with $a_\mathrm{s} \approx 0.1$ nm and $N_\mathrm{a} \approx 4000$.

17. A. A. Sukhorukov *et al.*, Phys. Rev. Lett. **92**, 093901 (2004).

18. L. Dobrek *et al.,* Phys. Rev. A **60**, R3381 (1999).

19. L. D. Carr and Y. Castin, Phys. Rev. A **66**, 063602 (2002).

20. E. S. Polzik *et al.*, Phil. Trans. Roy. Soc. **361**, 1391 (2003).

# Figure captions

Figure 1 (color online). Splitting of a three-soliton bound state when $\alpha_{\text{in}} = 0.1$ (a), 0.3 (b), 0.5 (c), 1.4 (d). Lattice depth $p = 0.05$ and modulation period $T = \pi/2$.

Figure 2 (color online). Propagation angles of solitons arising upon splitting of bound states of two (a) and three (b) solitons as a function of the incident angle. In (a) and (b) lattice depth $p = 0.05$. (c) Field distribution at the distance $\xi = 50$ for soliton with amplitude $\chi_1$ arising upon decay of the two-soliton bound state when $\alpha_{\text{in}} = 0.2$, $p = 0.05$ (black curve) and $p = 0.045$ (red curve). Black curve in (d) corresponds to that in (c), while red curve shows field distribution for soliton arising upon decay of bound state at $\alpha_{\text{in}} = 0.2$, $p = 0.05$ in the presence of white noise with variance $\sigma_{\text{noise}}^2 = 0.01$. Arrows in (c) and (d) show shift direction for a soliton maximum position. Modulation period $T = \pi/2$.

Figure 3 (color online). Splitting of the three-soliton bound state when $\alpha_{\text{in}} = 0.1$ and $p = 0.02$, (a) and $p = 0.2$ (b). (c) Propagation angles of solitons arising upon splitting of a three-soliton bound state versus lattice depth at $\alpha_{\text{in}} = 0.1$. The soliton with amplitude $\chi_5$ always stay in the central lattice channel. Modulation period $T = \pi/2$.

Figure 4 (color online). Localization of the single solitons produced by the splitting of a three-soliton bound state when $\alpha_{\text{in}} = 0.08$ (a) and 0.66 (b). Lattice depth $p = 0.6$ and modulation period $T = \pi/2$.

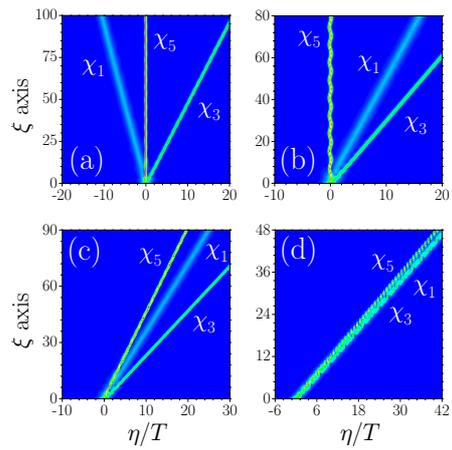

Figure 1 (color online). Splitting of a three-soliton bound state when $\alpha_{\text{in}} = 0.1$ (a), 0.3 (b), 0.5 (c), 1.4 (d). Lattice depth $p = 0.05$ and modulation period $T = \pi/2$.

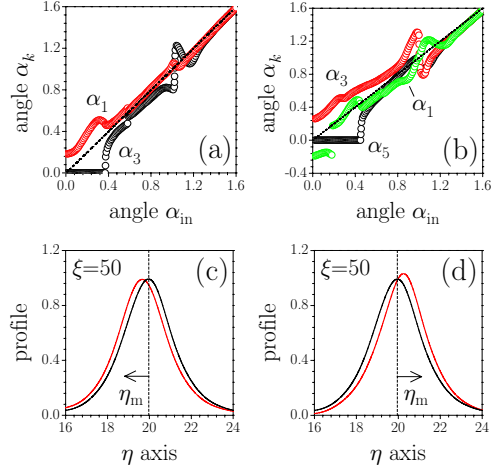

Figure 2 (color online). Propagation angles of solitons arising upon splitting of bound states of two (a) and three (b) solitons as a function of the incident angle. In (a) and (b) lattice depth $p=0.05$. (c) Field distribution at the distance $\xi=50$ for soliton with amplitude $\chi_1$ arising upon decay of the two-soliton bound state when $\alpha_{\rm in}=0.2$, $p=0.05$ (black curve) and $p=0.045$ (red curve). Black curve in (d) corresponds to that in (c), while red curve shows field distribution for soliton arising upon decay of bound state at $\alpha_{\rm in}=0.2$, $p=0.05$ in the presence of white noise with variance $\sigma^2_{\rm noise}=0.01$. Arrows in (c) and (d) show shift direction for a soliton maximum position. Modulation period $T=\pi/2$.

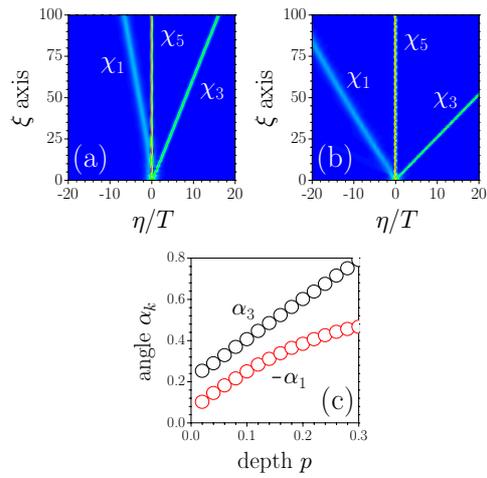

Figure 3 (color online). Splitting of the three-soliton bound state when $\alpha_{\rm in} = 0.1$ and $p = 0.02$, (a) and $p = 0.2$ (b). (c) Propagation angles of solitons arising upon splitting of a three-soliton bound state versus lattice depth at $\alpha_{\rm in} = 0.1$. The soliton with amplitude $\chi_5$ always stay in the central lattice channel. Modulation period $T = \pi/2$.

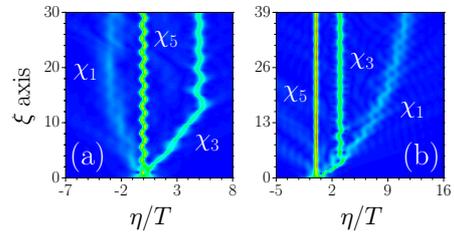

Figure 4 (color online). Localization of the single solitons produced by the splitting of a three-soliton bound state when $\alpha_{\text{in}} = 0.08$ (a) and 0.66 (b). Lattice depth $p = 0.6$ and modulation period $T = \pi/2$.